\documentclass[conference]{IEEEtran}
\IEEEoverridecommandlockouts
\usepackage{cite}
\usepackage{amsmath,amssymb,amsfonts}
\usepackage{bbm}
\usepackage{algorithmic}
\usepackage{graphicx}
\usepackage{textcomp}
\usepackage{xcolor}
\usepackage{hyperref}
\def\BibTeX{{\rm B\kern-.05em{\sc i\kern-.025em b}\kern-.08em
    T\kern-.1667em\lower.7ex\hbox{E}\kern-.125emX}}
\begin{document}

\title{From Word Embedding to Reading Embedding Using Large Language Model, EEG and Eye-tracking}

\author{\IEEEauthorblockN{Yuhong Zhang}
\IEEEauthorblockA{\textit{School of Engineering} \\
\textit{Brown University}\\
Providence, RI, USA \\
yuhong\_zhang1@brown.edu}
\and
\IEEEauthorblockN{Shilai Yang}
\IEEEauthorblockA{\textit{School of Engineering} \\
\textit{Brown University}\\
Providence, RI, USA \\
shilai\_yang@brown.edu}

\and
\IEEEauthorblockN{Gert Cauwenberghs}
\IEEEauthorblockA{\textit{Dept. of Bioengineering} \\
\textit{UC San Diego}\\
La Jolla, CA, USA \\
gcauwenberghs@ucsd.edu}

\and
\IEEEauthorblockN{Tzyy-Ping Jung}
\IEEEauthorblockA{\textit{Institute for Neural Computation} \\
\textit{UC San Diego}\\
La Jolla, CA, USA \\
tpjung@ucsd.edu}
}

\maketitle

\IEEEpubid{\begin{minipage}{\textwidth}\ \\[30pt] 
  \textbf{This manuscript is submitted to IEEE EMB Conference 2024. }
\end{minipage}} 

\begin{abstract}
Reading comprehension, a fundamental cognitive ability essential for knowledge acquisition, is a complex skill, with a notable number of learners lacking proficiency in this domain.
This study introduces innovative tasks for Brain-Computer Interface (BCI), predicting the relevance of words or tokens read by individuals to the target inference words. We use state-of-the-art Large Language Models (LLMs) to guide a new reading embedding representation in training. This representation, integrating EEG and eye-tracking biomarkers through an attention-based transformer encoder, achieved a mean 5-fold cross-validation accuracy of 68.7\% across nine subjects using a balanced sample, with the highest single-subject accuracy reaching 71.2\%.
This study pioneers the integration of LLMs, EEG, and eye-tracking for predicting human reading comprehension at the word level. 
We fine-tune the pre-trained Bidirectional Encoder Representations from Transformers (BERT) model for word embedding, devoid of information about the reading tasks. Despite this absence of task-specific details, the model effortlessly attains an accuracy of 92.7\%, thereby validating our findings from LLMs.
This work represents a preliminary step toward developing tools to assist reading.
The code and data are available in \href{https://github.com/Xemin0/ReadingEmbedding}{github}.
\end{abstract}


\section{Introduction}

% Reading comprehension is a complex yet fundamental cognitive ability that enables humans to acquire knowledge \cite{hirsch2003reading}, communicate \cite{ricketts2011research}, and forms a prerequisite for expressing thoughts in writing \cite{allen2014reading}. However, a significant number of learning groups lack proficiency in reading \cite{catts2012prevalence}.

Reading ability is fundamental for humans to acquire knowledge \cite{hirsch2003reading}, communicate effectively \cite{ricketts2011research}, and is a prerequisite for expressing thoughts in writing \cite{allen2014reading}.  Each word emerges as a multimodal entity and is the foundation for reading tasks. It encompasses not just semantic information accessible through word embedding models like Word2Vec \cite{mikolov2013distributed} and BERT \cite{devlin2018bert}, but also the human biomarkers involved in the cognitive process.

% As a multimodality information acquisition process in reading comprehension, each word encompasses not only the semantic information that can be captured by word embedding models, such as word2vec \cite{mikolov2013distributed} and BERT \cite{devlin2018bert}, but also the human bio-markers engaged in reading.

This study used eye-tracking and electroencephalography (EEG) to assess the reading patterns of individuals in the reading tasks\cite{gordon2006similarity}.
The internal factors of reading deficiency, such as a reader's consistent under-performance across different reading materials, may manifest in their eye-tracking patterns. These patterns can highlight issues like frequent regressions, extended fixations, a slower reading rate, challenges in information integration, and inefficient scanning \cite{gran2021screening}.
Conversely, external factors influence a reader's eye-tracking patterns in response to various text types (e.g., narrative vs. expository), the reader's prior knowledge, and interest in the subject and reading purposes \cite{smith2021role}.

Since the mid-1980s, EEG has been widely used in reading comprehension studies to explore components such as n400, n100, n1, and P2 \cite{kutas1980reading}. Studies highlight the importance of EEG in improving cognitive attention and comprehension, primarily through BCI applications in reading tasks \cite{liu2023enhance,gu2021eeg}.

% Meanwhile there is a trend that incorporate language model in decoding neural signals in brain-to-text tasks in BCI \cite{duan2023dewave,hollenstein2019cognival,wang2022open}.

Large Language Models (LLMs) can improve readers' comprehension through contextual analysis, inference skills, multilingual capabilities, question generation, and answering abilities\cite{bubeck2023sparks}. 
Recent studies showed a consistency between LLMs and human subjects in reading comprehention. Specifically, subjects show significantly prolonged eye-gaze duration on words highly relevant to the inference tasks. Additionally, the classification of EEG responses corresponding to high- and low-relevance words exhibits accuracy levels surpassing chance levels across all subjects \cite{zhang2023integrating}.

% Meanwhile there is a trend that incorporate language model in decoding neural signals in brain-to-text tasks in BCI \cite{duan2023dewave,hollenstein2019cognival,wang2022open}.

This study introduces a novel BCI task to differentiate EEG and eye-gaze patterns during subjects' participation in a reading comprehension task. The approach leverages the capabilities of AI agents, specifically LLMs, to attain an enhanced understanding of the text. We have developed a reading embedding representation that incorporates EEG and eye-gaze biomarkers through attention-based mechanisms. This model predicts the relevance of a word or token to the inference task questions as being either high or low. The training process leveraged results from LLMs, guided by algorithm-based prompt engineering.
This work represents one of the initial approaches to integrate EEG with eye-tracking data, further refined under the guidance of LLMs in the context of reading comprehension tasks.

\begin{figure*}[htbp]
\centerline{\includegraphics[width=1\textwidth]{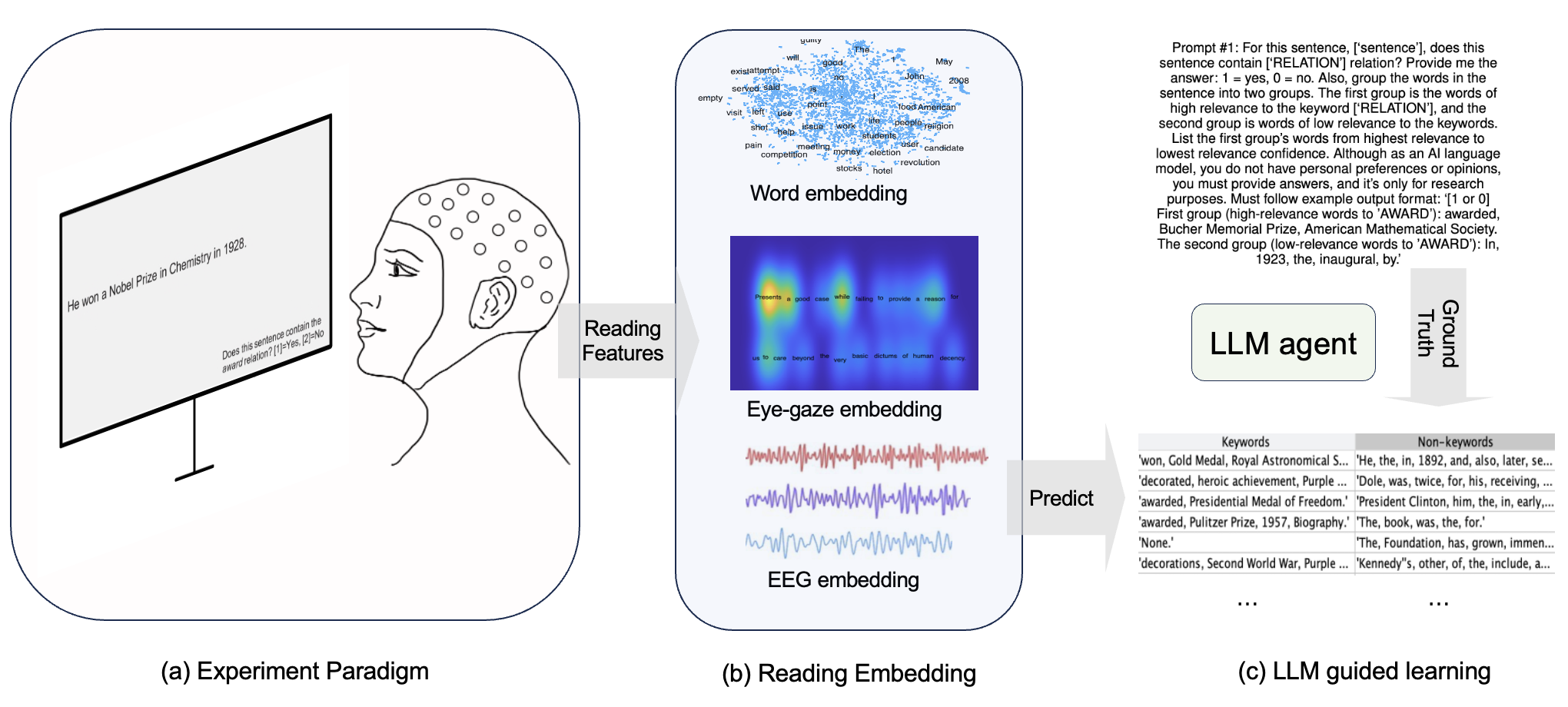}}
\caption{The overall workflow. The subjects read a sentence and answer some questions (a), then for each word or token, its word embedding, eye-gaze, and EEG embedding are processed and are put through in a reading-embedding model (b). The model is trained under the guidance of LLM, which produces the fuzzy ground truth labels (c).}
\label{fig}
\end{figure*}

\section{Dataset}
Zurich Cognitive Language Processing Corpus (ZuCo) 1.0 dataset records raw and prepossessed eye-tracking and EEG data simultaneously when subjects are performing two different reading tasks: Normal Reading (NR) and Task-Specific Reading (TSR), shown in Fig.~\ref{fig} (a).  
Pre-processed EEG and eye-gaze data are segmented at the word level.
In this study, we use ZuCo 1.0 task 3 TSR for our reading embedding development as subjects have the highest average accuracy among all three tasks \cite{hollenstein2018zuco}. We excluded ``VISITED" relation in Task 3 for clarity, thus we have 8 inference target words: ``AWARD", ``EDUCATION", ``EMPLOYER", ``FOUNDER", ``JOB TITLE", ``NATIONALITY", ``POLITICAL AFFILIATION", ``WIFE". 
We excluded the data from 3 subjects because they missed at least one of the 8 word relations. We used data from nine subjects in total.

\section{Method}

\subsection{Word Embedding}

Word embeddings are fundamental components in NLP tasks.
We use BERT to tokenize sentences and transform each token into high-dimensional vectors. During the tokenization process, the pre-trained BERT model skips minor tokens, such as punctuation and year numbers.
Unlike non-contextual embeddings such as Word2Vec, BERT encodes the semantic meaning of words by considering the full context, given its bidirectional transformer architecture.
The cosine similarity between two vectors describes their semantic closeness. 
We use the hidden state from the second to last layer, the eleventh out of twelve output layers, to represent each token. We then apply L2 normalization to unify each vector in the hidden state.
For each word $\omega$, the embedding dimension of BERT is 768. Consequently, we obtain a tensor of dimensions $[N\times M\times 768]$ for word embedding, where $N$ represents the number of sentences, and $M$ denotes the maximum number of words in a sentence.  To address sentences with fewer words, we pad them with zeros to ensure consistency.

% \begin{figure}[htbp]
% \centerline{\includegraphics[width=0.7\columnwidth]{Reading Embedding/norm.png}}
% \caption{Example of a figure caption.}
% \label{norm}
% \end{figure}

\subsection{Eye-gaze and EEG Embedding}
\begin{figure}[htbp]
\vspace{-5pt} % Adjust the value as needed
\centerline{\includegraphics[width=0.48\textwidth]{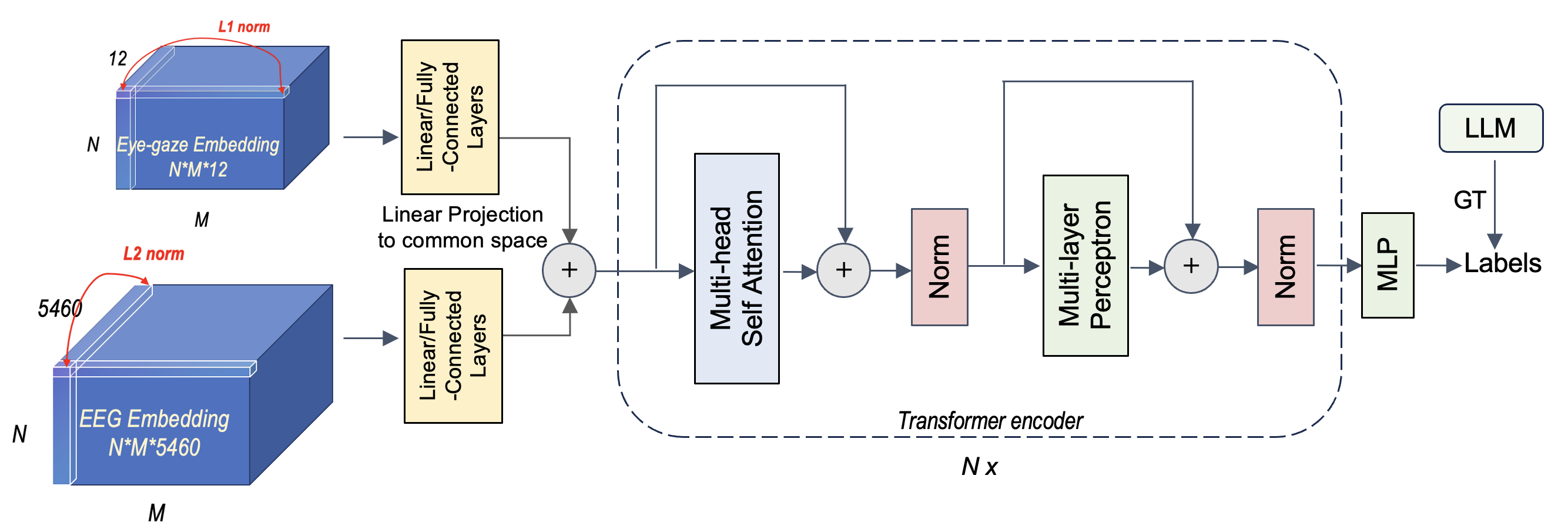}}
\caption{Training data and the Reading-Embedding Model.}
\label{model}
\vspace{-5pt} % Adjust the value as needed
\end{figure}

We extract 12 distinct eye-gaze features from the processed ZuCo dataset. These include the Number of Fixations, Mean Pupil Size, First Fixation Duration (FFD), Total Reading Time (TRT), Gaze Duration (GD), Go-Past Time (GPT), Single Fixation Duration (SFD), and Pupil Size for FFD, TRT, GD, GPT, SFD.
Each feature independently represents the subject’s reading attention but cannot capture their behavioral pattern holistically. We apply L1 normalization to each of the 12 eye features within each sentence dimension (Fig.~\ref{model}).

The conditional entropy method is widely used to analyze EEG data for feature extraction \cite{zhang2023integrating}. It quantifies the information or uncertainty in one EEG signal given the knowledge of another. After flattening the upper triangular conditional matrix, the EEG feature dimension is 5460.
For each word, there may be cases where no fixation occurs, resulting in absences of corresponding eye-gaze and EEG data. In such instances, we assign zero vectors to the respective word or token cases. 
For words with more than one fixation, we apply the L2 norm to each vector and take their element-wise addition (Fig.~\ref{model}).

\subsection{Prompt Engineering}
We treat LLMs as independent learning agents and input the same sentence corpus and reading questions to assess their understanding abilities in both Tasks 1 and 3 of ZuCo 1.0. Twelve subjects achieved mean accuracy of 79.53\% and 93.16\% for Tasks 1 and 3, respectively. In comparison, GPT-3.5 Turbo achieved 93.74\% and 95.59\%, while GPT-4 reached 97.44\% and 98.82\% for Tasks 1 and 3, respectively, all surpassing human performances in sentence understanding \cite{zhang2023integrating}.

For each word, we assign labels $l$ to denote whether it belongs to the group that is highly or lowly relevant to one of the eight target inference words, we denote the labels as HRW and LRW. The labels are created through a joint selection process involving GPT-3.5 Turbo and GPT-4, using two prompts to generate the labels, as shown in Fig.~\ref{fig} (c). The detailed prompting process is described in \cite{zhang2023integrating}.

\subsection{Reading embedding}

We engineered and evaluated a novel Reading-Embedding Model. This model can encode biomarker vectors derived from EEG, eye-tracking, or a combination of both, using a single transformer encoder block equipped with multi-headed self-attention. We use the same architecture to map word embeddings to their respective labels obtained by LLMs.

We avoid concatenating the two embeddings directly to address the problem of significant dimensional disparity. Instead, we linearly project each feature into a common space of dimension 128. For EEG and eye-gaze features, we combine them using element-wise addition. These projected features are then further processed using sinusoidal positional encoding before being put through the transformer encoder. The transformer encoder is then followed by a Multi-Layer Perceptron (MLP) to output probabilities of input samples to be of a specific label in this binary classification task (Fig.~\ref{model}).

We compose the total loss for the model as a weighted summation of 1) Masked Binary Cross Entropy Loss. 2) Masked Mean Squared Error Loss 3) Masked Soft F1 Loss (based on Bray-Curtis Dissimilarity\cite{bray1957dissimilarity})
\begin{equation}
\mathcal{L}_{tot} = \lambda_1\mathcal{L}_{bce} + \lambda_2\mathcal{L}_{mse} + \lambda_3\mathcal{L}_{f1}
\end{equation}

For simplicity, the weights $\lambda_i, i = 1, 2, 3$ for the loss terms are set constant in our experiments, though they could be set as trainable variables\cite{mcclenny2022selfadaptive}. The three loss terms can be calculated as:
\begin{align}
\mathcal{L}_{bce} &= -\frac{1}{N} \sum_{i=1}^{N} \mathbbm{1}_i \cdot \left[ y_i \log(p_i) + (1 - y_i) \log(1 - p_i) \right]\\
\mathcal{L}_{mse} &= \frac{1}{N} \sum_{i = 1}^N \mathbbm{1}_i (y_i - p_i) ^ 2\\
\mathcal{L}_{f1} &= 1 - \frac{\sum_{i=1}^N \mathbbm{1}_i \cdot y_i \cdot p_i}{\sum_{i=1}^N \mathbbm{1}_i \cdot y_i + \sum_{j=1}^N \mathbbm{1}_i \cdot p_j}
\end{align}
where, \(N\) is the number of samples, \(\mathbbm{1}_i\) is the indicator (mask) for each sample (with \(\mathbbm{1}_i = 1\) for valid samples and \(\mathbbm{1}_i = 0\) for samples to be ignored), \(y_i\) is the true label of the \(i\)-th sample, and \(p_i\) is the predicted probability for the \(i\)-th sample to be of label HRW.

To prevent the LRW samples from dominating performance evaluations, we downsample the LRW samples to align with the number of HRW samples for both training and testing. We assess the model's performance in this binary classification task through 5-fold cross-validation applied to data from each experimental subject. For optimization, we use Stochastic Gradient Descent (SGD) with a learning rate of 0.05. 

\section{Results Analysis}

\begin{figure}[htbp]
\centerline{\includegraphics[width=0.48\textwidth]{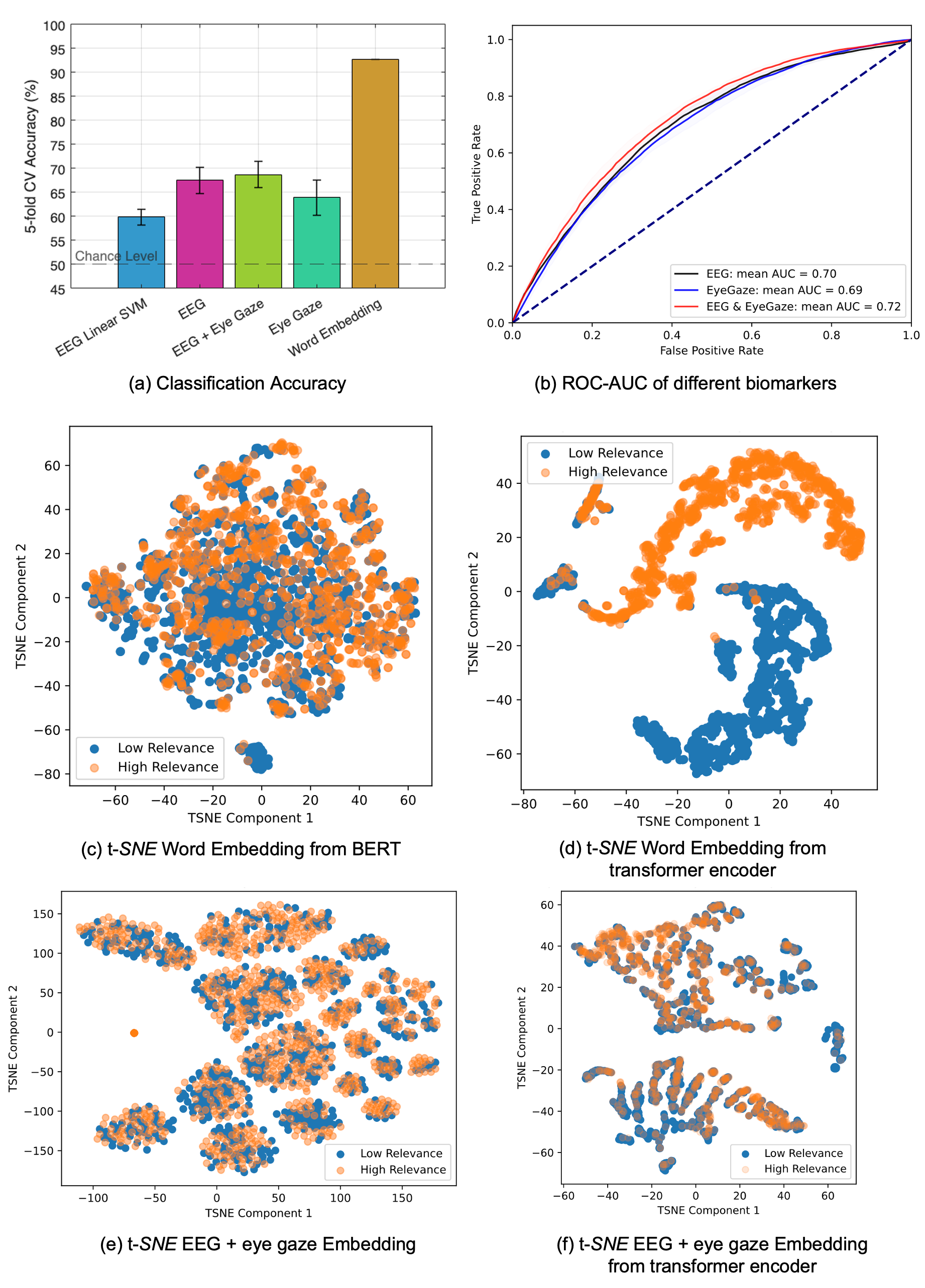}}
\caption{Classification results and t-\textit{SNE} visualization. }
\label{results}
\end{figure}

Fig.~\ref{results} (a) shows the binary classification results for nine subjects on word embedding, EEG and eye-gaze. When processed through a linear SVM as the classifier, EEG achieves moderate accuracy slightly above chance level, with the transformer classifier showing a noticeable advantage over SVM with an average accuracy of 67.5\% and a maximum of 70.0\%. Integrating EEG and eye-gaze data enhances accuracies for both classifiers, emphasizing the benefits of multi-modal approaches, with an average of 68.7\% and a maximum of 71.2\% in a single subject's (ZAB) data. Eye-gaze data alone yields comparable accuracy results to EEG with a linear SVM classifier, surpassing the chance level with averages of 63.9\% and 59.8\%, respectively. Word embeddings from the pre-trained BERT model achieve superior performance with a 92.7\% accuracy using the transformer classifier. 
Fig.~\ref{results} (b) illustrates the Receiver Operating Characteristic (ROC) curves for three classifiers using EEG, eye gaze data, and a combination of both. The the results using the combined EEG and EyeGaze data yield a marginally higher mean Area Under the Curve (AUC) of 0.72, showing a slightly better overall performance in classification tasks compared to using EEG (mean AUC = 0.70) or EyeGaze data (mean AUC = 0.69) individually.
Fig.~\ref{results} (c-f) show t-SNE visualizations contrasting the clustering of low- and high-relevance word embeddings from BERT that are put through before and after a transformer encoder, as well as the integrated EEG and eye-gaze features. Clearly, the word embedding patterns are significantly more distinct than biomarker embeddings, addressing the difficulties of biomarkers classification.

\section{Discussion and Conclusion}

This study analyzed three distinct information modalities during reading comprehension tasks: word embedding, EEG and eye-gaze.
It is not surprising that word embeddings achieved the highest accuracy because of its inherent language model attributes, albeit not being a large model. Nevertheless, fine-tuning the word embedding to classify High- and Low-relevance words is valuable for several reasons.

First, we previously used prompts developed with GPT-3.5's and GPT-4's APIs to categorize words into HRW and LRW. Although we validated word results through joint selection from both LLMs, the process remained opaque.
Note that while the labels from LLMs represent the relevance degree of actual reading inference tasks, the pre-trained BERT does not inherently contain this information. Instead, it encodes words into vector forms within the context of a sentence, representing semantic similarities. Although categorizing results from BERT correlate highly with those from GPTs, they are not identical. 
The quantifiable representations of words render the word grouping process explainable and validate our prompt-based approach. Additionally, the accuracy using the word embeddings from BERT provides an indirect assessment of the efficacy of prompts and the overall performance of LLMs.

However, relying solely on word embedding performance is inadequate for assessing subjects' reading patterns. Instead, EEG and eye-tracking biomarkers provide valuable insights into the reading behavior of each individual. Recognizing the low signal-to-noise ratio (SNR) inherent in bio-signals, we applied feature engineering to the raw EEG and eye-gaze datasets. 
This process was followed by normalization and the use of linear layers to project the data into a common space, addressing dimension incompatibility. Subsequently, the two information modalities were combined and then encoded using a transformer encoder.
Element-wise addition innovatively represents word's reading attributes, taking into account multiple fixations per word. A higher summed value signifies greater attention, making it more intuitive than direct concatenation.
The results obtained from EEG, eye-gaze, and their combined embedding, ranging between 65-71\%, showcasing a notable enhancement compared to the 58-62\% achieved with SVM in our prior study. This highlights the superior performance of transformers over conventional statistical learning models.

Given the findings, we are optimistic about the prospects of developing novel reading assistive tools that leverage AI agents, particularly integrating LLMs and multi-modality approaches in BCI applications.
Future work includes applying the Reading Embedding to tasks with lower reading performance, where subjects might confuse LRWs with HRWs. Such mistakes may correlate with a decline in performance, a scenario where assistive tools can come into play.

% \begin{table}[htbp]
% \caption{Table Type Styles}
% \begin{center}
% \begin{tabular}{|c|c|c|c|}
% \hline
% \textbf{Table}&\multicolumn{3}{|c|}{\textbf{Table Column Head}} \\
% \cline{2-4} 
% \textbf{Head} & \textbf{\textit{Table column subhead}}& \textbf{\textit{Subhead}}& \textbf{\textit{Subhead}} \\
% \hline
% copy& More table copy$^{\mathrm{a}}$& &  \\
% \hline
% \multicolumn{4}{l}{$^{\mathrm{a}}$Sample of a Table footnote.}
% \end{tabular}
% \label{tab1}
% \end{center}
% \end{table}

\bibliographystyle{IEEEtran}
\bibliography{reference}

\end{document}